\documentclass[showpacs,preprintnumbers,amsmath,amssymb,twocolumn]{revtex4}

\usepackage{graphicx}

\newcommand{\expectation}[1] {\left\langle #1 \right\rangle}

\newcommand{\vol}[1] {{\bf #1}}

\begin{document}

\preprint{preprint number}

\title{\bf One Dimensional Chain with Long Range Hopping }

\author{Chenggang Zhou and R. N. Bhatt }

\affiliation{ Dept of Electrical Engineering \\ Princeton University,
  Princeton NJ 08544 }

\date{\today}

\begin{abstract}

The one-dimensional (1D) tight binding model with random nearest
neighbor hopping is known to have a singularity of the density of
states and of the localization length at the band center. We study
numerically the effects of random long range (power-law) hopping with
an ensemble averaged magnitude $\expectation{|t_{ij}|} \propto
|i-j|^{-\sigma}$ in the 1D chain, while maintaining the particle-hole
symmetry present in the nearest neighbor model. We find, in agreement
with results of position space renormalization group techniques
applied to the random XY spin chain with power-law interactions, that
there is a change of behavior when the power-law exponent $\sigma$
becomes smaller than 2.  \end{abstract}

\pacs{PACS number }

\maketitle

\section{Introduction}
\label{Introduction}

The one-dimensional (1D) tight-binding model of non-interacting
electrons with random nearest neighbor hopping has been extensively
studied since Dyson's exact solution to density of states of
one-dimensional random harmonic oscillators \cite{dyson}, which can be
mapped to the random nearest neighbor hopping model.  Mertsching gave
a node counting scheme to study the density of states in a similar
model\cite{mertsching}. Since the early 80s, supersymmetry methods
have been used to study such problems with
randomness\cite{efetov,pafisher}.  All these studies show that there
is a singularity in the one-electron density of states
$\rho(\epsilon)$ at the band center of the form $\rho(\epsilon) \sim
{1 \over \epsilon |\ln\epsilon|^3}$.  Concurrently, as a consequence
of the Thouless theorem \cite{thouless}, there is a logarithmic
divergence of the localization length $ \xi \sim |ln \epsilon
|$. Thus, this model exhibits a critical point at the band center,
unlike the standard Anderson model\cite{anderson} in one-dimension
with on-site disorder, where all states are known to be
localized\cite{mott} for any finite disorder. This is a consequence of
the particle-hole symmetry that is maintained in the model despite
randomness in the hopping when only nearest neighbor hopping is
present.\\

        In this paper, we investigate in whether these features change
if the nearest neighbor hopping model is generalized to include long
range hopping. Since these properties are closely related to the
particle-antiparticle symmetry in the Hamiltonian, we constrain the
long range hopping to a form which preserves the particle-hole
symmetry. (Operationally, this is done by allowing hopping only
between sites separated by an odd number of lattice constants.)
However, the long range hopping terms render analytic methods such as
node counting theorom \cite{mertsching} and Thouless
theorem\cite{thouless} inapplicable; consequently we perform numerical
calculations to study this problem instead.\\

The nearest neighbor tight binding model of non-interacting fermions
can be mapped onto a XY spin chain with nearest neighbor random
couplings via a Jordan-Wigner transformation\cite{lieb}.  (See, {\em
e.g.} McKenzie\cite{rmackenzie} for a recent discussion of the
topic). The singularity in the density of states of the fermion
problem at zero energy translates to a singular magnetic
susceptibility at zero temperature in the spin problem.
Fisher\cite{dsfisher}, in particular, showed how a perturbative
position space renormalization group (RG) scheme using the ratio of
small to large couplings for random spin systems employed by several
groups in the 80s in the context of random quantum spin-1/2
chains\cite{dasgupta, bhattlee} becomes asymptotically exact, and
yields the correct low temperature (energy) behavior. Thus the correct
form of the singularity at the band center is obtained for the
fermionic model from the asymptotically exact RG. The same position
space RG technique has been used recently to study the XY spin chain
with long range (power-law) exchange $ J(r) \propto r^{-\sigma} $
\cite{houck}.  The behavior found there suggests that the strong
disorder fixed point, which determines the functional form of low
temperature magnetic susceptibility in the spin problem, (for nearest
neighbor hopping this is related to the density of states singularity
in the fermionic model) survives for power law exponents $\sigma$
above a critical value, which they determined to be 2 (within their
numerical accuracy of 5-10 per cent ).

Using the Jordan Wigner transformation on the 1D XY spin chain with
long range couplings, while yielding a particle-hole symmetric fermion
Hamiltonian, does not give a pure noninteracting fermion problem with
power law hopping [ terms such as $c^\dagger_i exp(i\pi
\sum_{n=i}^{j-1} { c^\dagger_n c_n })c_j $ appear as well ].  While
the {\em exact} connection between the fermionic and spin models is
lost when further neighbor hoppings are added to the Hamiltonian, we
can nevertheless expect some similarities between the two models - XY
random spin chain and free fermions with 1D random hopping - when the
two have exchanges/hoppings with a power-law fall off, at least for
large power-law exponents. The result for the long range spin model
also motivates paying close attention to our model's behavior when
$\sigma$ is in the vicinity of 2.

The outline of our paper is as follows.  In the next section, we
define the model Hamiltonian. The following section discusses the
methods and relevant formalisms used in computation. We then present
the results for the density of states, localization length of
eigenstates, and the spin-spin correlation function for the equivalent
magnetic model, obtained from various numerical calculations. We find
that the density of states retains the singularity at the band center,
but this singularity is gradually reduced by the long range
hopping. The eigenstates are not well described as exponentially
localized states, but have power-law-like tails determined by
$\sigma$. Finally, the spin-spin correlation function also shows
power-law decay where $\sigma$ is not too small.

\section {The model}
\label{The model}

  The model we have studied is built on a 1D bipartite lattice,
consisting of odd and even sites along the chain. The Hamiltonian
contains fermionic hopping terms between the two sublattices
consisting of the odd and even sites, respectively, while hopping
within the same sublattice is forbidden. Thus we have,

\begin{eqnarray}
    H = \sum_{<ij>} { t_{ij} (c_i^\dagger c_j + c_j^\dagger c_i) }
\end{eqnarray}

where $i$ and $j$ are summed over members of the odd and even
sublattices respectively.  This bipartite property preserves the
particle-antiparticle symmetry and leads to interesting properties in
the density of states and localization length. The hopping parameters
$t_{ij}$ are random with zero average, defined in terms of the
distance $|i-j|$ and a Gaussian random variable, as described below.
Two models with power law fall-off of $\expectation{|t_{ij}|}$ have
been studied.  Model A allows hopping between every pair formed by
picking one site from each of the two sublattices, while $<|t_{ij}|>$
decreases with distance as a power law for each pair
$(i,j)$. Specifically, we take

\begin{eqnarray*}
 t_{ij} = { w_{ij} \over \left|i-j\right|^\sigma }
\end{eqnarray*}

Where $w_{ij}$ is a Gaussian random number with distribution
\begin{eqnarray*}
P(w_{ij})= {1 \over a \sqrt{2\pi}} \exp \left(-{w^2_{ij} \over 2
a^2}\right)
\end{eqnarray*}

 In model B, $<|t_{ij}|>$ is constant, but the probability for a
hopping term to be present decreases with the hopping distance as a
power-law. Thus

 \begin{eqnarray*} t_{ij} = w_{ij}M(\left|i-j\right|) \end{eqnarray*}

where $w_{ij}$ is a Gaussian random variable as in Model A, and
$M(\left|i-j\right|)$ is chosen randomly to be either 1 (with
probability $\left|i-j\right|^{-\sigma} $) or 0 [with probability
$(1-\left|i-j\right|^{-\sigma} $)].

We expect these two models to exhibit some similarities. We have used
a = 0.2 throughout this paper. In both models, therefore, we have only
one parameter $\sigma$ which characterizes the model. A small value of
$\sigma$ produces longer range hopping.\\

\section{Computational Details}

The Hamiltonian of this model is an $N \times N$ matrix for a system
of N sites, with elements $t_{ij} = 0$ when $|i-j|$ is a multiple of
2. Because of this bipartite property, the Schr\"{o}dinger equation
can be written as two coupled equations when number of sites $N$ is
even. Let $\psi^{odd}_i$ be the wavefunction on the sublattice
consisting of the odd numbered sites, and $\psi^{even}_i$ that on the
sublattice consisting of even sites. If the matrix $h$ represents the
hopping terms from even sites to odd sites, then $h^\dagger$ will
consist of the hopping terms from odd sites to even sites.  The
Schr\"{o}dinger equation can then be written as two equations:
\begin{eqnarray}
  h\psi^{even} &=& \epsilon \,\psi^{odd} \\ h^\dagger \psi^{odd} &=&
  \epsilon \, \psi^{even} .
\end{eqnarray}

These equations lead to

\begin{eqnarray}
\label{e2}
h^\dagger h \psi^{even} = \epsilon^2 \, \psi^{even}\\
\label{e3}
h h^\dagger \psi^{odd} = \epsilon^2 \, \psi^{odd}
\end{eqnarray}

The above equations reduce the size of the matrix of the $ N \times N$
Hamiltonian to $N/2 \times N/2$. Further, from the above equations, it
is clear that the eigenvalues come in pairs $\pm \epsilon$, and their
eigenvectors are related by

\begin{eqnarray}
\label{e5}
  \psi^{even} _\epsilon &=& \psi^{even} _ {-\epsilon} \\
\label{e6}
  \psi^{odd}_\epsilon &=& -\psi^{odd}_{-\epsilon}
\end{eqnarray}

A direct consequence of the above is that the density of states is
symmetric for each realization of the disorder. By exploiting the
particle-antiparticle symmetry, we gain efficiency in our numerical
calculation. \\

The density of states can be obtained by directly diagonalizing large
matrices. In this approach, the difficulty is that we have to use
large matrices which do not have significant finite size effects.
Since we're interested in the property of the singularity at the band
center, which by particle-hole symmetry, lies at $ \epsilon = 0 $, we
need to get enough statistics near the band center.  We identify
finite size effects by comparing the results of different sizes.\\

The other approach to calculating the density of states is a recursive
method. First, the dense matrix is cut off at a range large enough, so
that the remaining part is an acceptable approximation for the
original matrix of arbitary size.  [ For finite range hopping, this
method gives no approximation at this stage, but for long-range (power
law hopping) this is a different cutoff scheme, with a somewhat
different finite size effect ]. Then the matrix $H-\epsilon$ is
transformed into diagonal form by a similarity transformation, which
rotates certain columns and rows to eliminate off-diagonal terms. The
remaining diagonal elements are not eigenvalues, but they retain the
signature of the matrix, {\em i.e.,} a positive element implies an
eigenvalue satisfying $E_i>\epsilon$, a negative element corresponds
to an eigenvalue satisfying $E_i<\epsilon$. By counting the number of
positive or negative diagonal elements we can obtain the integrated
density of states. This process can be continued for arbitary length
with no finite-size effect as in direct diagonalization, until the
statistical error of the integrated density of states is smaller than
our requirement.  For nearest neighbor model, the recursion equation
for diagonal element is:
\[ \zeta_n = -\epsilon - {H_{nn-1}^2 \over \zeta_{n-1}}\]
where $H_{nn-1}$ is the only off-diagonal term in the Hamiltonian,
$\zeta_n$ is the remaining diagonal element after the transformation.
Suppose $\zeta_n$ has a distribution $F(\zeta)$ as $n \longrightarrow
\infty$, then it is given by the integral equation

\[ F(\zeta) = \int_{-\infty}^{+\infty}
        \delta(\zeta+\epsilon+{x \over \xi}) P(x) F(\xi) dx d\xi \]

where $P(x)$ is the distribution function of $H_{nn-1}^2$. Although
this equation is similar to Dyson's approach \cite{dyson}, it has not
been solved analytically due to the peculiar argument of the $\delta$
function. The numerical approach is just to generate a sequence of
$\zeta_n$ and calculate its distribution.\\

To take care of the finite size effect introduced by the cut-off, we
compare results using different cut offs. Such a procedure allows us
to ascertain the range of energies for which the integrated density of
states has converged: lower the energy, longer the cut-off required
(the cutoff required varies roughly logarithmically with energy). For
a fixed cutoff, the energy variation of the integrated density of
states obtained by this method mimics that of the nearest neighbor
{\em i.e.,} finite range model - this sets the lower bound of energies
close to the band center for which this method is applicable for that
cutoff. \\

Once we have obtained the eigenfunctions by direct diagonalization, we
can calculate the correlation function of the corresponding spin
model. The Jordan-Wigner transformation \cite{lieb} transforms the
XY-spin chain in zero external field to a half-filled band of
fermions, because the following identity

\begin{eqnarray}
  \sum_i S_i^z = \sum_i c_i^\dagger c_i - {N/2}
\end{eqnarray}

The ground state fills the $N/2$ states with negative energy which we
can get by diagonalizing the Hamiltonian. We can further calculate the
spin correlation function on the ground state. We compute the zz part
of the spin-spin correlation function defined by

\begin{eqnarray}
  C(i,j) = \overline{\expectation{S^z_iS^z_j}}
\end{eqnarray}

$\expectation{...}$ denotes the expectation value in the ground state,
while the bar on top denotes ensemble average over the disorder. With
$S^z_i = c^\dagger_i c_i - {1 \over 2}$, and $i \neq j$, the
correlation function can be written as

 \begin{eqnarray} C(i,j) = \overline{\expectation{c^\dagger_j
  c^\dagger_i c_i c_j}}-{ 1 \over 4}
\end{eqnarray}

where we have used the fact $\overline{\expectation{c^\dagger_i c_i}}
= {1 \over 2}$ due to half filling.  The Jordan-Wigner transformation
contains a phase factor due to the anticommutation relation of fermion
operators on different sites, but the phase factor coming from
different fermion operators cancels each other in our expression of
$C(i,j)$.  In the random singlet phase\cite{pafisher} appropriate for
describing the nearest neighbor only model in the spin operator
language, all these correlation functions have the same dependence on
distance.  The four fermion term can be expanded into a Hartree term
and a Fock term.

\begin{eqnarray*}
  \expectation{c^\dagger_j c^\dagger_i c_i c_j} = \sum_{n,m,occupied}
  \psi_n^\dagger(i) \psi_n^\dagger(i) \psi_m(j) \psi_m(j)\\
  -\psi_n^\dagger(i) \psi_n^\dagger(j) \psi_m(j) \psi_m(i))
\end{eqnarray*}

Here the sum is over all occupied states, i.e. all N/2 states with
negative energy. The Hartree term can be evaluated directly to be $1 /
4$ because the wavefunctions are orthonormal using equations
(\ref{e5}) and (\ref{e6}).  Therefore the correlation function is
given by the Fock term only.  Further, if $i$ and $j$ are on the same
sublattice, we see from equations (\ref{e2}) and (\ref{e3}) that
$\psi_n(i)$ and $\psi_n(j)$ ( n is among $N/2$ occupied states) are
both eigenvectors of either $h^\dagger h$ or $ hh^\dagger$, and
consequently orthogonal to each other. Thus, when $i$ and $j$ are on
the same sublattice, the correlation function is exactly zero. In the
numerical results presented below, we only display the spin
correlation functions between 2 sublattices, and $C(i,j)$ is replaced
by $C(x)$ where $2x+1 = |i-j|$.
\begin{eqnarray}
  \label{eq:1} C(x) = -\sum_{m,n,occupied} \overline{
   \psi_n^\dagger(i) \psi_n^\dagger(j) \psi_m(j) \psi_m(i) }
\end{eqnarray}

This expression can be evaluated directly, but since the computing
time for evaluating C(x) is even more than that required for
diagonalization, the system sizes we use are smaller than for
diagonalization. The data presented for C(x) are all obtained from
systems of 256 sites. However, we are able to discern the behavior
reasonably well from the data for $C(x)$ subject to this
limitation. \\

\section{Density of States}
\label{density of states}

\begin{figure}
\includegraphics[angle=-90,width=3.5in]{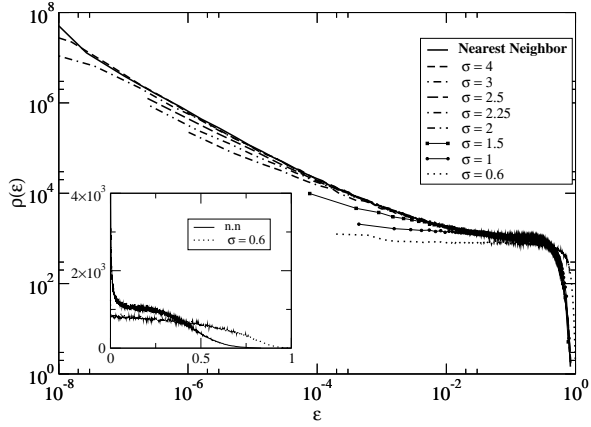}
\caption{\label{fig:1} Density of states of model A for different
values of the power law exponent $\sigma$, shown on a double
logarithmic plot, showing the singularity at the band center
($\epsilon = 0$). The inset shows the data on a linear scale.  The
singularity becomes very weak when $\sigma$ is less than 2, and is not
discernable for the sizes studied when $\sigma$ is below 1.  }
\end{figure}

\begin{figure}
\includegraphics[angle=-90,width=3.5in]{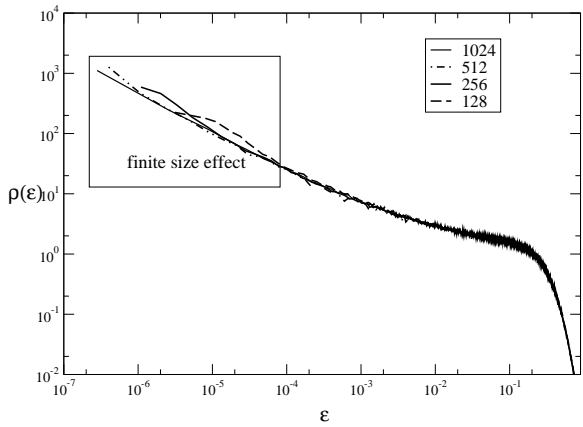}
\caption{\label{fig:2} The same data as in figure \ref{fig:1}, plotted
in a different way as motivated in the text.  As expected, the density
of states of the nearest neighbor model asymptotically falls on a
straight line. The upwards bending of the curves with decreasing
$\sigma$, evident for $\sigma < 3$ shows the weakening of the density
of states singularity by the long range hopping.}
\end{figure}

Our results on model A show that the density of states remains
singular at the band center when the long range hoppings are present.
Figure \ref{fig:1} shows the density of states $\rho(\epsilon)$
obtained by diagonalization as a function of energy $\epsilon$ on a
double logarithmic plot for both the nearest neighbor model, and for
the long range model for different values of the power law exponent
$\sigma$. As can be seen, the singularity at the band center
($\epsilon = 0$) persists at least for large $\sigma$, though its
magnitude clearly decreases. The inset to Fig \ref{fig:1} compares the
data for the nearest neighbor model along with the data for the lowest
power law ($\sigma = 0.6 $) on a linear scale, which gives a clearer
idea of the extent of this decrease. For quantitative purposes, it is
better to plot the same data as $(\rho(\epsilon)\epsilon)^{1/3}$
vs.$\ln\epsilon$.  On such a plot, shown in fig \ref{fig:2}, the
nearest neighbor model is supposed to lie on a straight line, which it
clearly does.  Further, the curves show little deviation from the
nearest neighbor model as long as $\sigma > 3 $. The deviation for
smaller $\sigma$ are consistent with the singularity being gradually
weakened as $\sigma$ decreases; if we fit the data with
$\rho(\epsilon) \sim {1 \over \epsilon |\ln\epsilon|^\omega}$ then
when $\sigma =2$, the best fit $\omega$ is about 5.  As $\sigma$
becomes lower than 1, direct diagonalization is almost incapable of
revealing any detail of the singularity: we only see a large value at
the band center, and the density of states approaches the Wigner
semicircle\cite{wigner}. Nevertheless, the thermodynamic limit remains
well defined upto $\sigma = 0.5$, below which the bandwidth starts to
increase with system size, i.e. we need to scale the hopping magnitude
with system size to have a sensible thermodynamic limit. This critical
value of $\sigma$ can be predicted by writing the model in
path-integral form, using either replica technique or supersymmetry,
and averaging over the random variables. Such an approach gives a four
fermion term proportional to $\Sigma |i-j|^{-2\sigma}$, where $i$ and
$j$ are site indices, therefore when $\sigma$ is less than 0.5, the
sum will diverge. \\

\begin{figure}
\includegraphics[angle=-90,width=3.5in]{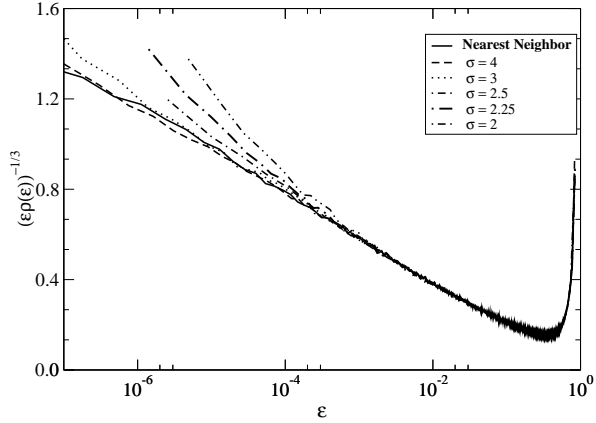}
\caption{ \label{fig:3} Finite size effects on density of states,
which appear at the low energy end of these curves. Data are density
of states for four different size systems with the same $\sigma =
2.5$, with length varying from $N = 128$ to $1024$.}
\end{figure}

We have diagonalized several different sizes of samples to ensure that
the data shown are not corrupted by finite size effects.
Fig. \ref{fig:3} shows an example of a finite size effect, which
appears as a size-dependent rounding of density of states at low
energy. All the data plotted correspond to N = 1024 sites unless
stated otherwise, and do not suffer significant finite size effects.
The results of model B are very similar to model A, except that the
smallest $\sigma$ for a proper thermodynamic limit to exist is 1.\\

\begin{figure}
\includegraphics[angle=-90,width=3.5in]{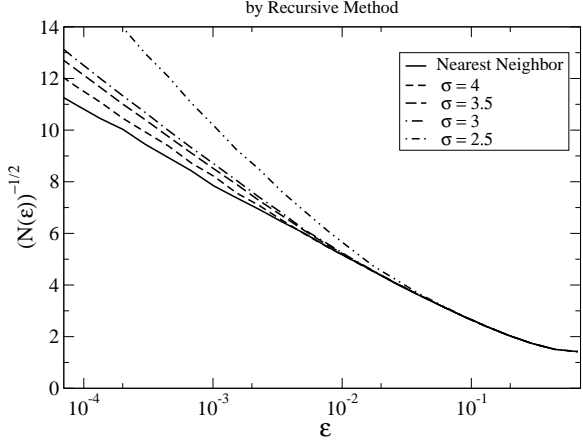}
\caption{ Integrated density of states, computed by the recursive
method, with cut-off at 100th neighbor. Within the displayed energy
range, these curves are checked to be free of effects due to the
finite cut-off in the hopping range. Note that the curves are smoother
than the results obtained from straight diagonalization. The nearest
neighbor model is expected to asymptotically fall on a straight line
with a slope equal to -1. Curves fitted by the formula $N(\epsilon)
\sim {1/|\ln\epsilon|^v}$, yield $v$ somewhat larger than the value 2
appropriate for the nearest neighbor model, for $\sigma$ below 3. }
\label{fig:4}
\end{figure}

Fig \ref{fig:4} shows our results for the density of states obtained
using the recursive method. Here, one obtains the integrated density
of states $ N(\epsilon)$ (from $0$ to $\epsilon$).  For the nearest
neighbor model, the exact asymptotic form is given by:
\[N(\epsilon) \sim {1 \over |\ln\epsilon |^2}\]
Fig \ref{fig:4} plots the inverse square root of $N(\epsilon)$ versus
$ \ln \epsilon$; for the nearest neighbor model, the expected straight
line behavior is seen. For power law hopping, the data shows
measurable curvature certainly for $\sigma = 2.5$.  For larger
$\sigma$ it is difficult to see whether the data suggest curvature, or
simply a changing slope with decreasing $\sigma$. While it is tempting
to fit these curves with a form like
\[N(\epsilon) \sim  |\ln\epsilon |^ {- \lambda}\] ,
which will lead to an singularity in the density of states like $
\rho(\epsilon) \sim 1 / ( \epsilon |\ln\epsilon |^{\lambda+1} )$, the
data are better fit with several $\lambda$. This suggests that
corrections to the asymptotic form may be important for power law
hopping.\\

\begin{figure}
\includegraphics[angle=-90,width=3.5in]{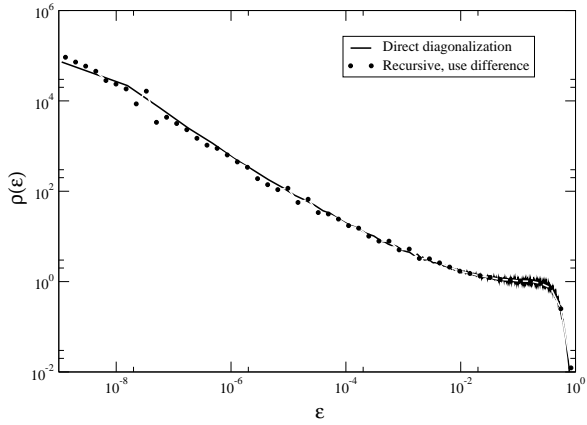}
\caption{\label{fig:5} Comparison between the recursive method and the
direct diagonalization method for the density of states.  The solid
curve is by direct diagonalization, while the points are density of
states obtained by taking the difference of integrated density of
states from the recursive method. The agreement between the curves is
good, and no systematic errors are found.}
\end{figure}

\begin{figure}[h]
\includegraphics[angle=-90,width=3.5in]{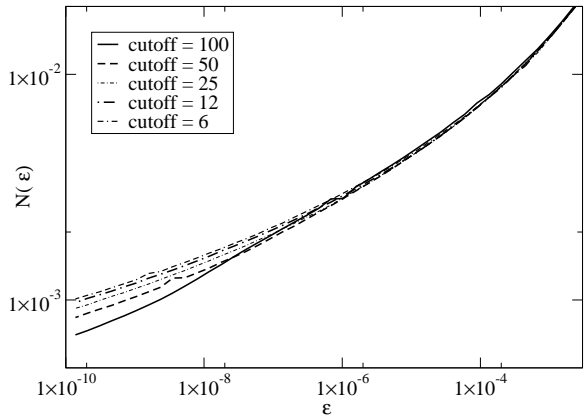}
 \caption{Convergence of the recursive method exhibited using
 different cutoff parameters for the hopping.  Small values of the
 cutoff lead to deviations starting from higher energies. }
\label{fig:6}
\end{figure}

In figures \ref{fig:5},\ref{fig:6}, we show the accuracy of this
recursive method. Figure \ref{fig:5} shows a direct comparison of the
two methods.The recursive method agrees with the diagonalization
results before the finite size effect sets in. Figure \ref{fig:6}
shows the convergence of the recursive method as the range of cut-off
is increased.  (Below a certain energy, which corresponds to distances
beyond the cutoff length, the recursive method will behave like
nearest model, since the wavefunction spreads out of the range of
hopping).


The density of states allows us calculate the specific heat and spin
susceptibility for this this model of non-interacting fermions.  Using
the standard formulae\cite{smith}, we can see that the specific heat
prefactor ($\gamma$, where $c_v = \gamma T$) and susceptibility
($\chi$) at low temperature are singular at the center of the
band. Thus, in the vicinity of the band center the zero temperature
susceptibility is the form:
\[ \chi \sim { 1 \over T|\ln T|^{\omega-1}} \]

if the singularity in density of states is $ 1 \over \epsilon
|\ln\epsilon|^{\omega}$. A similar formula holds for $\gamma$.

\section{Localization Length}
\label{Localization Length}

The nearest neighbor model is known to have a sigularity in the
localization length ($ \xi $) at center of the band. It's asymptotic
form is
\begin{eqnarray}
 \xi \sim |\ln \, \epsilon |
\end{eqnarray}

This singularity can be deduced from the singularity of density of
states using the Thouless theorem\cite{thouless}. However, in the
longe range hopping model, the theorem does not apply, and our
numerical calculation suggests that the behavior of the two models is
rather different.\\

For finite energy, all states are found to decay from a central
maximum in both models, and the decay becomes weaker as the band
center is approached, as in the nearest neighbor model. However, the
behavior with distance is rather different for the two models.
\begin{itemize}
\item
In the case of model A which has genuine long range (power law)
behavior of the hopping parameter $t_{ij}$, the tail of the
wavefunctions actually decay in a power-law manner, $\psi(x) \sim
x^{-\sigma}$. This form is obviously determined by the power-law long
range hopping term. If we apply the usual method of looking at the
asymptotic behavior to determine localization length, the localization
length is infinite for any power law exponent!
\item

In the case of model B, the wavefunction is found to be decaying
exponentially at long distances, like the nearest neighbor model, and
the localization length can be obtained by several methods, which
agree with the theoretical prediction.
\end{itemize}

\begin{figure}[h]
\includegraphics[angle=-90,width=3.5in]{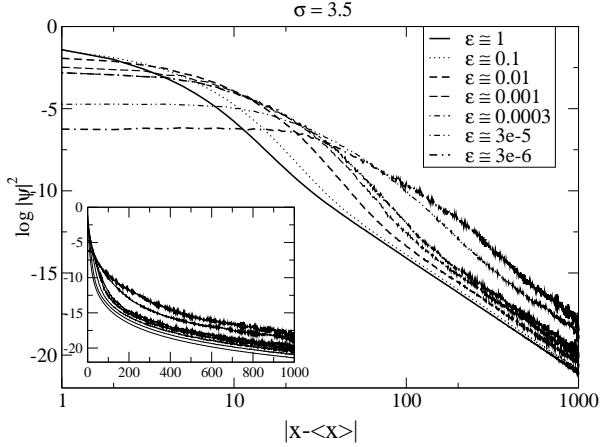}

  \caption{Double logarithmic plot of the probability density in real
space for individual eigenstates plotted from the center of each
state, averaged over eigenstates with a given energy $\epsilon$ for
model A with $\sigma = 3.5$. Small values of $\epsilon$ correspond to
states near the band center.  We can see as the energy approches the
band center, the states become more delocalized. In the tail, for all
energies, the profile of the wavefunction actually decreases in a
power-law fashion, with an exponent exactly determined by the value of
$\sigma$. Similar plots for other values of $\sigma$ in model A show
the same feature. The inset with the probability density on a semilog
plot, shows distinct curvature in the tail, unlike an exponentially
localized state.}  \label{fig:7}
\end{figure}

Fig \ref{fig:7} shows a log-log plot of the averaged probability
density (wavefunction amplitude squared $|\psi|^2$) as a function of
the distance from the center of the wavefunction, averaged over
typically 512 states, as a function of energy away from the band
center for model A. At long distances, the behavior is clearly linear
on this double logarithmic plot, implying a power law decay at long
distances. Fitting the data shows clearly that the decay is related to
the power law behavior of $t_{ij}$ - the tail of $|\psi|^2$ decays as
$ 1 \over |x-<x>|^{-2\sigma}$.  This tells us that we cannot use the
localization length, as defined by Thouless for an exponentially
decaying state here. However, we may usefully define moments of the
wavefunction to compute, for example, inverse participation
ratios\cite{ipr}.\\

\begin{figure}[h]
\includegraphics[angle=-90,width=3.5in]{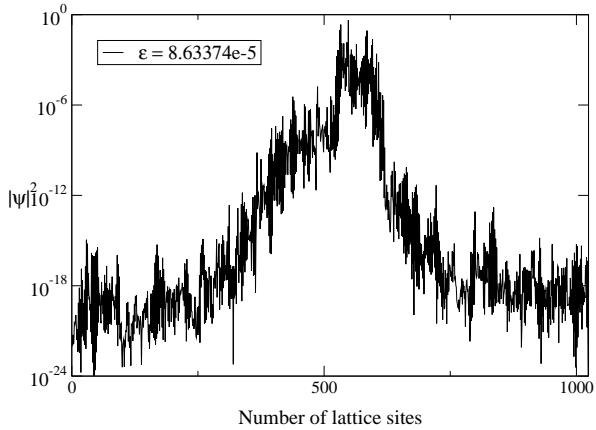} \caption{A typical
        wavefunction of model A near the band center(small $\epsilon$)
        shown on a log-linear plot}
\label{fig:8}
\end{figure}

\begin{figure}[h]
\includegraphics[angle=-90,width=3.5in]{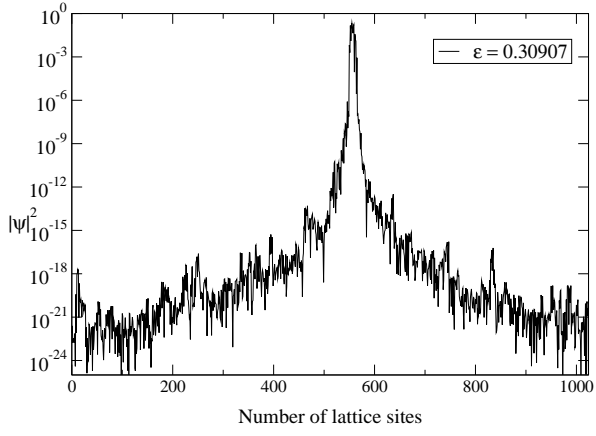}
\caption{A typical wavefunction of model A away from the band
center(larger $\epsilon$) on a log-linear plot.  The tail shows clear
curvature on this plot, implying a functional form that decays slower
than exponential.}  \label{fig:9}
\end{figure}

\begin{figure}[htbp]
\includegraphics[angle=-90,width=3.5in]{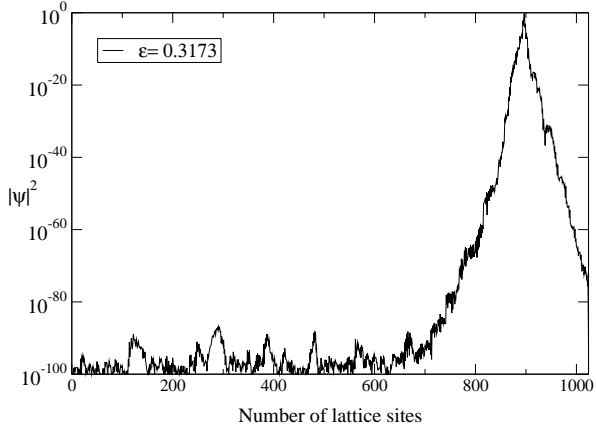} \caption{A typical
        wavefunction of model B away from the band center. A clear
        exponential decay of the wavefunction amplitude over 50
        decades is seen, as in the nearest neighbor model.}
        \label{fig:10}
\end{figure}

\begin{figure}[h]
\includegraphics[angle=-90,width=3.5in]{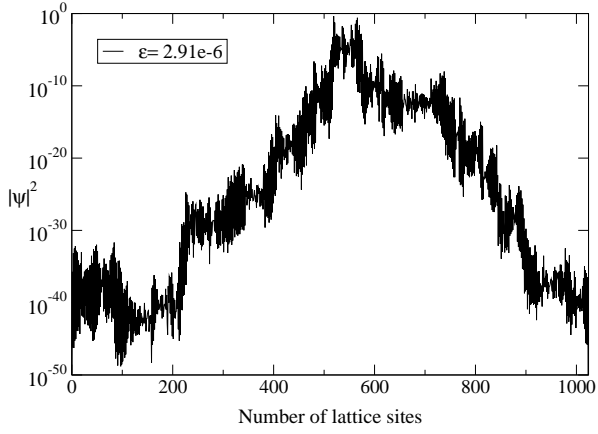} \caption{A typical
        wavefunction of model B near the band center. Despite strong
        fluctuations, the main profile is basically exponential, as in
        the nearest neighbor model.}  \label{fig:11}
\end{figure}

The problem of defining the localization length in model A is not
shared by model B, where we find that the wavefunctions always decay
exponentially. To show this difference, in Figures \ref{fig:8} to
\ref{fig:11} we have plotted typical wavefunctions for model A and
model B ({\em on a log-linear plot of probability density versus
distance}) near the center and in the tail of the band. The difference
between the two models near the center of the band is not obvious due
to the large fluctuations, but in the tail of band, they look clearly
different - model B shows straight exponential decay (see
Fig. \ref{fig:10}) down to 50 orders of magnitude for $|\psi|$, while
model A (see Fig. \ref{fig:9} ) shows clear upwards concave curvature
in the tail, characteristic of a slower decaying function (like a
power law) at long distances.

\section{Spin Correlation Function}
\label{Correlation}

We now present results for the spin-spin correlation function for the
associated model in terms of the spin operators obtained by a Jordan
Wigner transformation. The motivation is to compare the results with
those obtained for the long range random antiferromagnetic XY spin
chain\cite{houck}.  We reiterate that because of long range hopping,
our model contains terms in addition to those in the pure power law XY
model studied by Houck and Bhatt; however, because of the existance of
the same long range two-spin interactions in both studies, there may
be several points in common.

\begin{figure}[h]
\includegraphics[angle=-90,width=3.5in]{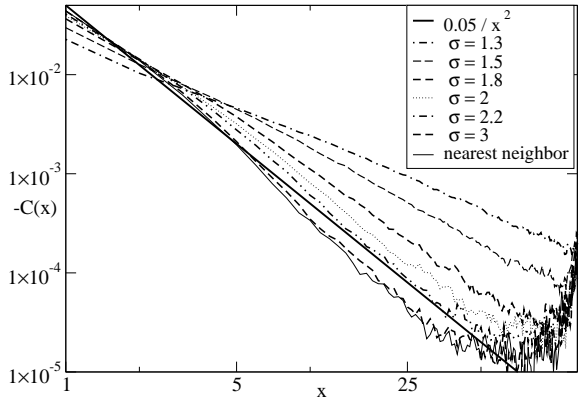}
\caption{Spin-spin correlation function in the ground state ($T=0$),
for the corresponding spin model obtained using a Jordan-Wigner
transformation. The bold straight line, which decays as $1 \over x^2$,
is a guide for the eye. We see that when $\sigma$ is above 2,
including nearest neighbor model, $C(x)$ exhibits this inverse square
behavior. Below $ \sigma = 2$, $C(x)$ is better fitted by $1 \over
x^\sigma$. More sample averaging is necessary to smooth the noisy tail
at long distances.}  \label{fig:12}
\end{figure}

\begin{figure}[h]
\includegraphics[angle=-90,width=3.5in]{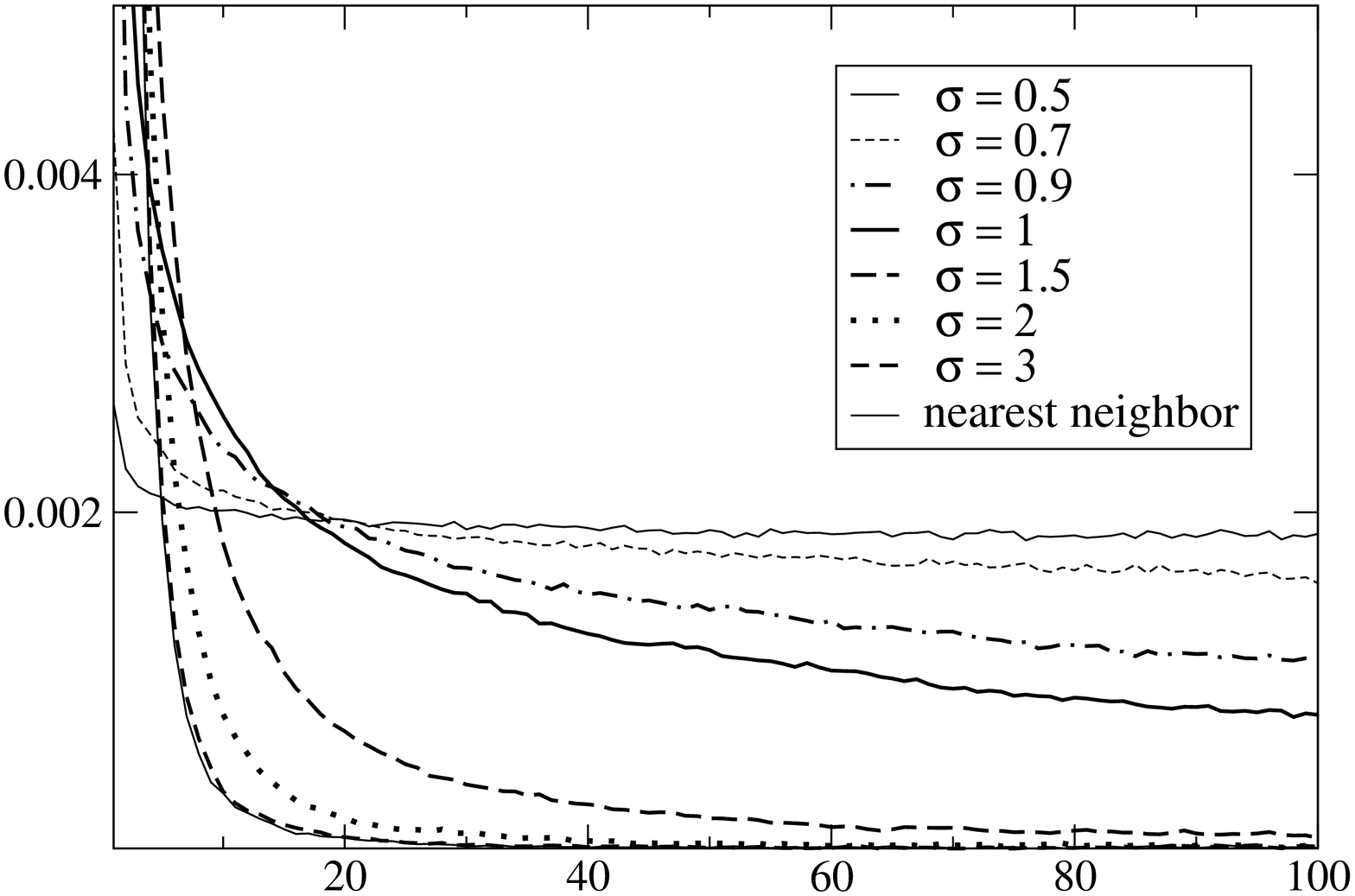}
\caption{Spin-spin correlation function, as in figure \ref{fig:12},
plotted on a linear scale. For $ \sigma < 1$ $C(x)$ decays very
slowly, and seems to approach a constant value at large distances.}
\label{fig:13}
\end{figure}

Figure \ref{fig:12} shows the average correlation function $C(x)$
obtained by averaging 256 samples of model A each having 256 sites
plotted on a double logarithmic plot. All the values of $C(x)$ are
negative (corresponding to antiferromagnetic correlations). Our
results suggest that there are three regions of distinct behavior,
which can be summarized as given below:

\begin{itemize}
\item
For fast power law decays ({\em i.e.,} exponents $\sigma > 2$ ), the
long distance behavior of the correlation function remains unchanged
from the nearest neighbor model. Thus in Figure \ref{fig:12}, the
curves are parallel to each other at large $x$ values, consistent with
the result for the nearest neighbor model, for which a slope of 2 is
predicted\cite{dsfisher}.  (For example, a best fit of $C(x)$ for x
within the interval $[20,60]$ $\sigma=2.2$ yields a slope of 1.97).

\item
When $\sigma$ gets below 2, the slope begins to change. For $\sigma$
close to but less than 2, the slope appears to be given by $\sigma$
itself, implying $C(x) \sim x^{- \sigma} $.  ( A least squares fit to
$C(x)$ of the form $ C(x) = { a \over x^v } $, gives $v=1.79$ for
$\sigma = 1.8$, $v = 1.44 $ for $\sigma = 1.5$ and $v = 1.12$ for
$\sigma = 1.3$).  Our numerically exact results thus show that the
model's magnetic counterpart changes its low energy behavior different
behavior for power law exponents $\sigma$ below 2. This is precisely
the value around which Houck and Bhatt\cite{houck} found that the
perturbative real-space RG procedure appears to break down, perhaps
signalling a change of phase.
\item
Below $\sigma \sim 1$, the deviations of $C(x)$ from the nearest
neighbor model become rather significant. To exhibit this more
clearly, we show a linear plot in Figure \ref{fig:13}.  C(x) seems to
decay very slowly, and our data are consistent with it approaching a
finite (negative) value, implying antiferromagnetic order. (For
example, the curve for $\sigma = 0.7$ is well fitted by ${a \over
x^{v}}+d$, where $ v = 0.49 $ is 1/2 within numerical accuracy). We
caution, however, that for such low power law decays, finite size
effects can be large, and more detailed calculations with larger
system sizes and more samples is necessary before this result can be
stated with certainty from numerical studies.
\end{itemize}

In summary, the numerical results on the spin correlation function
from the long range fermion model (which we can solve numerically
exactly) supports the earlier observations on the XY chain with random
long range couplings using perturbative numerical RG methods that the
random singlet phase is unstable for power law couplings with
exponents less than 2.  However, unlike the numerical RG study, which
sees this as a breakdown of the RG scheme, we are able to go into the
new phase, which appears to be characterized by continuously varying
exponents, like a critical phase.  We also find evidence for a
possible transition to long range order at still smaller $\sigma$. It
should, however, be borne in mind that these observations, are from
numerical calculations in {\em finite} systems, and subject to
statistical errors due to finite sampling of the quenched random
variable.

\section{Summary}
\label{Summary}
In this paper, we have presented results of a numerical study of a one
 dimensional lattice model of non-interacting fermions with random
 long range (power-law) hopping, which maintains particle-hole
 symmetry of the nearest neighbor model by allowing hopping only
 between even and odd sites ({\em i.e.,} no hopping allowed between
 odd sites, or between even sites).  We have studied two models - one
 with genuine power-law hopping, and another with long range hopping
 with a power law fall-off of the probability of such a hopping to be
 present. The results on density of states, localization and spin
 correlation function of the two models have been presented and
 analysed.

For the density of states, we observe that the singularity at the
center of the band, present for the nearest neighbor model, is
weakened by long range hopping (Model A).  The change is gradual, and
at least for power-law exponents $\sigma$ greater than 3, is
consistent with a change in the prefactor of the singularity. For
$\sigma$ less than 3, though, the numerical data for the desity of
states in the range available appear to fit better with a somewhat
different power of the logarithm of the energy.  Beyond $\sigma = 1$,
the data are consistent with no singularity being present, until
$\sigma<0.5$, when the thermodynamic limit becomes
ill-defined. Similar results are seen in Model B, except that the
thermodynamic limit becomes ill-defined at $\sigma = 1$.\\

The two models exhibit rather different behavior of the electronic
wavefunctions. Model B is conventional, in that its wavefunctions are
exponentially localized, just as the eigenstates of nearest neighbor
model. In model A, however, the wavefunctions are actually localized
in a power-law manner rather than exponential.  Consequently, the
usual definition of localization length in terms of the logarithm of
the long distance behavior of the wavefunction is invalid; however,
several inverse participation ratios can still be defined.

By transforming the fermion model back to a spin model using
Jordan-Wigner transformation, we calculated the spin correlation
function in the ground state.  Based on the data, three different
phases may be possible - (1) the random singlet phase, which seems to
be stable for power law exponents down to $\sigma = 2$; (2) a critical
type phase with a continuously varying exponent of the power-law
characterizing the spin-spin correlation function between $\sigma = 2$
and $ \sigma = 1$; and (3) a possibly long-range ordered phase for
$\sigma < 1 $.

\section{Acknowledgement}
\label{sec:ack}

This research was supported by NSF DMR-9809483.

\newpage

%
%
%


\begin{thebibliography}{99}
\bibitem{ipr} For example, the nth inverse participation ratio may be
defined by $ IPR(n) = \sum_i {|\psi(i)|^{2n+2}}$ where $\psi(i)$ is
the normalized wavefunction.
\end{thebibliography}
\end{document}